\documentclass[12pt]{iopart}

\usepackage{graphicx}
\usepackage{dcolumn}
\usepackage{bm}
\begin{document}

\title{Effect of a fluctuating parameter mismatch in coupled R\"{o}ssler systems}

\author{Manu P. John, Jijo P. Ulahannan and V. M Nandakumaran}

\address{International
School of Photonics, Cochin University of Science and
Technology.\\Cochin 682 022, India \\
email:manupj@cusat.ac.in}
\begin{abstract}
This paper is concerned with the effect of parameter fluctuations
with a characteristic waiting time in coupled R\"{o}ssler
oscillators. We show that the averaged error in synchronization that
is introduced due to a fluctuating parameter is proportional to the
waiting time and the amplitude of the fluctuations. It is also shown
that coupling strength beyond a threshold value does not have any
significant effect on improving the quality of synchronization when
the fluctuations posses considerable waiting time.
\end{abstract}

\maketitle

\maketitle
%\section{Introduction}
Synchronization of coupled chaotic systems has generated a lot of
research activities over the last several years
\cite{peco1,peco2,yamada}. Synchronized chaotic behavior has been
studied extensively in physical, chemical and biological systems.
One of the frequently employed methods is coupling two identical
systems, this coupling may be unidirectional or bidirectional.
Complete synchronization of identical chaotic systems is of
considerable interest because of its applications in secure
communication\cite{war1,roy2,war2,bindu}. By identical systems we
mean a set of systems whose parameters are exactly equal. It is
found that complete synchronization is not possible when there is a
small but finite mismatch of the parameters of the systems. In
coupled non autonomous systems effect of a phase mismatch or finite
constant frequency detuning is to destroy the synchronization
altogether\cite{yin}.

Coupled R\"{o}ssler oscillators have been extensively used for
studies in synchronization. This is due to their simplicity and
results obtained can usually be generalized to other chaotic
systems. Kurths. et. al have shown that coupled R\"{o}ssler systems
which possess parameter mismatch  attain a state of phase
synchronization\cite{phase} and on increasing the coupling strength,
such systems may synchronize, but lagged in time \cite{pslag}. It is
also found that many systems, both mathematical models and actual
physical systems, show similar behavior \cite{kurth}.  There remains
a question as to what happen if the parameter mismatch is
fluctuating. Such a study is relevant and important since in reality
it is difficult, if not impossible, to construct identical systems
except in numerical simulations. This can also be due to the fact
that the parameters could be fluctuating in time with time scales of
their own, either due to internal instabilities or due to some
external perturbations. In this paper we discuss the effect of
fluctuating parameters and the effect of timescales associated with
such fluctuations on the synchronization of coupled R\"{o}ssler
oscillators.
\section{Parameter fluctuations}
 \label{paraflt}
Let us consider the parameters $p_1$ and $p_2$ of a coupled system
of oscillators which fluctuates randomly. The fluctuations are
assumed to occur in time as follows
 \begin{eqnarray}
 \label{addfluct}
 p_{1t} = p_0+\xi_{1t}\\
 p_{2t} = p_0+\xi_{2t},\nonumber
 \end{eqnarray}
where, $\xi_{1t}$ and $\xi_{2t}$ are two delta correlated zero mean
random variables and $p_0$ be some average value of the parameters.
It is also assumed that the the fluctuations have a waiting time
$\tau_0$. That is $\tau_0$ is the time for which the parameter
retains its value before it is changed once again. We define the
fluctuation rate $\phi$, as the number of times a parameter is
perturbed in unit time which is also the inverse of the waiting time
. In a case where the parameter fluctuates in this fashion, it can
be seen the the parameter is correlated in time as
\begin{displaymath}
    \rho(p(t) p(t+\tau)) =\left\{
    \begin{array}{cc}
        1-\frac{\tau}{\tau_0}  & \mbox{if }\tau < \phi  \\
        \mbox{0} & \mbox{otherwise}
    \end{array}\right.
\end{displaymath}
Where $\rho(..)$ denotes the normalized auto correlation. The
amplitude of the parameter mismatch is denoted as $\widetilde{\Delta
p}$, given by
\begin{equation}
\widetilde{\Delta p}=\langle \mid \delta p (t) \mid \rangle_t,
\end{equation}
where, $ \delta p_{t} = p_{1t}-p_{2t}$ and $\langle ...\rangle_t$
denotes time average. We did not consider the root mean square value
because the system is sensitive to the magnitude of parameter
mismatch, not to its square, at least in general. This model of
fluctuation is, not the typical, but we believe that this is the
closest approximation that can be applied an actual system. Also, it
is interesting to note that our model of fluctuation is
complimentary to the dichotomic noise \cite{osc_ein} where amplitude
of fluctuation is a constant, but the waiting time fluctuates about
an average.

There is one question that may arise at this point. The systems
under consideration is continuous but the fluctuations are discrete.
Why continuous fluctuations, for example, an Ornstein Uhlenbeck
kind, which posses time scales and are continuous were not
considered. Our answer is, such fluctuations arise as a result of a
stochastic evolution. As far as complete synchronization is
considered, we do not expect such untethered stochastic evolution of
parameters in the parameter space. Instead, a parameter, if deviated
from its desired value is expected to fluctuate around an average as
complete synchronization is usually not found in nature but found in
fabricated systems, well monitored and well designed. There, we
expect if a parameter mismatch occurs, it will have a tendency or it
will be forced to cross zero very often. In such a situation the
prime consideration shall be the time scales associated with the
fluctuations.
\section{Effect of Fluctuations on Dynamics}
In a case where the parameter fluctuating as described in sec.
\ref{paraflt} the effect on the dynamics of coupled chaotic system
can be understood in terms of the dynamical equations. Let the
evolution of the coupled systems in phase space be given by,
\begin{eqnarray}
\label{dyana}
\dot{X_1}&=&f_1(p_1, X_1)+C f(X_2-X_1)\\
\dot{X_2}&=&f_1(p_2, X_2)+C f(X_1-X_2).\nonumber
\end{eqnarray}
Where $X$ represents the phase space variables, $p$ the parameter
whose fluctuation is considered, and $C$, the coupling strength.
With equation (\ref{dyana}) we can write an equation for the rate of
separation $ X_1- X_2 $ of the trajectories as,
\begin{equation}
\frac{d (X_1- X_2)}{dt}= \dot{X_1}- \dot{X_2}=M(p_1,p_2, X_1,X_2),
\end{equation}
 $M(p_1,p_2, X_1,X_2)$ is a function of the dynamical
variables, the parameters of the coupled systems and  $\Delta p$ the
parameter mismatch. This can be expanded int terms of $\Delta p$ and
the effect of fluctuations can be separated out.
\begin{eqnarray}
M(p_1,p_2, X_1,X_2)&=&M_s(p_0,X_1,X_2)\\
& &+E(X_1,X_2,\Delta p_1,\Delta p_2).\nonumber
\end{eqnarray}
Here $M_s(p_0,X_1,X_2)$ represents the quantity which offers a
stable synchronization manifold and $E(X_1,X_2,\Delta p_1,\Delta
p_2)$ represents the effect of the parameter mismatch.

Let us now see the form of $E(X_1,X_2,\Delta p_1,\Delta p_2)$ in the
in case of a system of bidirectionally coupled R\"{o}ssler
oscillators.
\begin{eqnarray}
\label{coup} \dot{x}_{1,2}&=&- y_{1,2} -
z_{1,2}+c(x_{2,1}-x_{1,2})\\\nonumber \dot{y}_{1,2}&=& x_{1,2} +
p_{1,2}y_{1,2}\\\nonumber \dot{z}_{1,2}&=& 0.2 + z_{1,2} ( x_{1,2} -
10 )\\\nonumber
\end{eqnarray}
Here it is assumed that in the absence of fluctuations and for an
appropriate value of $c$, the systems get synchronized. Also in the
presence of fluctuations an approximate synchrony is maintained due
to the negative conditional Lyapunov exponents and the zero mean
nature of the fluctuations. With equation (\ref{coup}), we can write
the rate of separation of trajectories as,
\begin{eqnarray}
\label{sperat2} \frac{d (x_1-x_2)}{dt} & =& -y_1+y_2-z_1+z_2+2 C
(x_2-x_1)\\\nonumber \frac{d (y_1-y_2)}{dt} & =&
x_1-x_2+p_1y_1-p_2y_2\\\nonumber \frac{d (z_1-z_2)}{dt}& = & x_1
z_1-x_2 z_2-10(z_1-z_2).\\\nonumber
\end{eqnarray}
Here it can be seen that the fluctuations affect the dynamics
through $y$, as the parameter $p$ occur only in the equation for
$\dot{y}$ in the coupled set of equations. Thus by Taylors expansion
around $p_0$, $E(X_1,X_2,\Delta p_1,\Delta p_2)$ can be written as,
\begin{eqnarray}\label{vanish}
E(X_1,X_2,\Delta p_1,\Delta p_2)= y_1 \xi_1-y_2\xi_2
\end{eqnarray}

The evolution of the system in time can be split into several sub
intervals of duration $\tau$. Thus in a approximately synchronized
state the error occurred due to the fluctuations in the $k^{th}$
interval can be written as,
\begin{eqnarray}
\label{accu_1} \epsilon(t, t+\tau)&=&\int_t^{t+\tau_0}
(y_{1}\xi_1-y_{2}\xi_2) dt
\end{eqnarray}
If the parameter fluctuation are fast enough $y_1$ and $y_2$ can be
considered to be a constant between two consecutive fluctuations and
is denoted by $\overline{y_1}$ and $\overline{y_2}$. Also $\xi_1$
and $\xi_1$ are constants for a given waiting time $\tau$ by
definition, and their value in the $k^{th}$ interval is $\xi_{1k}$
and $\xi_{2k}$. Thus equation (\ref{accu_1}) can be written as,
\begin{eqnarray}
\label{smapppush} \epsilon_k(t, t+\tau_0)&=&\int_t^{t+\tau}
(\overline{y_{1k}}\xi_{1k}-\overline{y_{2k}}\xi_{2k})dt\\\nonumber
&=& \tau_0(\overline{y_{1k}}\xi_{1k}-\overline{y_{2k}}\xi_{2k})
\end{eqnarray}
From this equation, $\langle \epsilon_k^2 \rangle$ can be calculated
\begin{eqnarray}
\langle \epsilon_k^2 \rangle & = & \tau_0^2 \langle
\overline{y_{1k}}^2 \xi_{1k}^2+\overline{y_{2k}}^2 \xi_{2k}^2
\rangle
\end{eqnarray}
In the above relation $\langle \overline{y_{1k}}^2
\xi_{1k}^2+\overline{y_{2k}}^2 \xi_{2k}^2 \rangle$ can be considered
as a constant as $y$ belong to an attractor which is dense in
periodic orbits and the averages of $\xi$ over the waiting time
follows some well defined statistical distribution. $\langle
\epsilon_k^2 \rangle_t$ can be written as
\begin{eqnarray}
\langle \epsilon_k^2 \rangle & = & \tau_0^2 \times constant
\end{eqnarray}
as $t~\rightarrow~\infty$. Here note that this expression do not
contain $C$ which is the coupling strength. This suggest that
increasing the coupling strength do not have any significant effect
on reducing the fluctuations. Also note that the error introduced by
fluctuations is proportional to the amplitude of fluctuations.
\section{Effect of Fluctuations on the Quality of Synchronization}
In the last section we saw that the the fluctuations affect the
dynamics in a multiplicative manner and its effect on the seperation
of the trajectories is proportional to the waiting time. In this
section we consider how the fluctuations affect the over all
dynamics of the coupled system. Consider a system where the dynamics
is represented by the dynamical equations $\dot{\nu_i}~=
f_i(\nu_1...\nu_n)$. The effect of a perturbation of a variable
$\nu_k$ on the variable $\nu_j$ can be written as $\delta
\nu_j=~\frac{\partial f_j(\nu_1...\nu_n)}{\partial \nu_k} \delta
\nu_k$. Thus it can be seen that in a situation where the phase
space variables can be considered to be a constant, the effect of
the perturbation applied to one variable have proportional effect on
other variables also.

Effect of such perturbations on the quality of synchronization can
be quantified using any measure of synchronization which is based on
the divergence of the trajectories of the coupled system. A well
known measure of the quality of synchronization which is the
similarity function $S(0)$ defined as,
\begin{equation}
\label{dynamic1}
S^2(0)=\frac{\langle[x_1(t)-x_2(t)]^2\rangle}{[\langle
x_1^2(t)\rangle \langle x_2^2(t)\rangle]^{\frac{1}{2}}}.
\end{equation}
In an ideally synchronized state this directly corresponds  to
$\epsilon_k$. If an approximate synchrony is maintained during the
evolution of the system we can assume that the majority of
contribution to the error in synchronization is of the form
$\epsilon_k$. Thus the analysis that we have presented is valid if
$S(0)$ can be fitted with a function which is linear in $\tau_0$, or
proportional to the inverse of $\phi$, the fluctuation rate in an
actual numerical experiment.
\subsection{Numerical simulations}
\begin{figure}
\includegraphics[width=0.9\columnwidth]{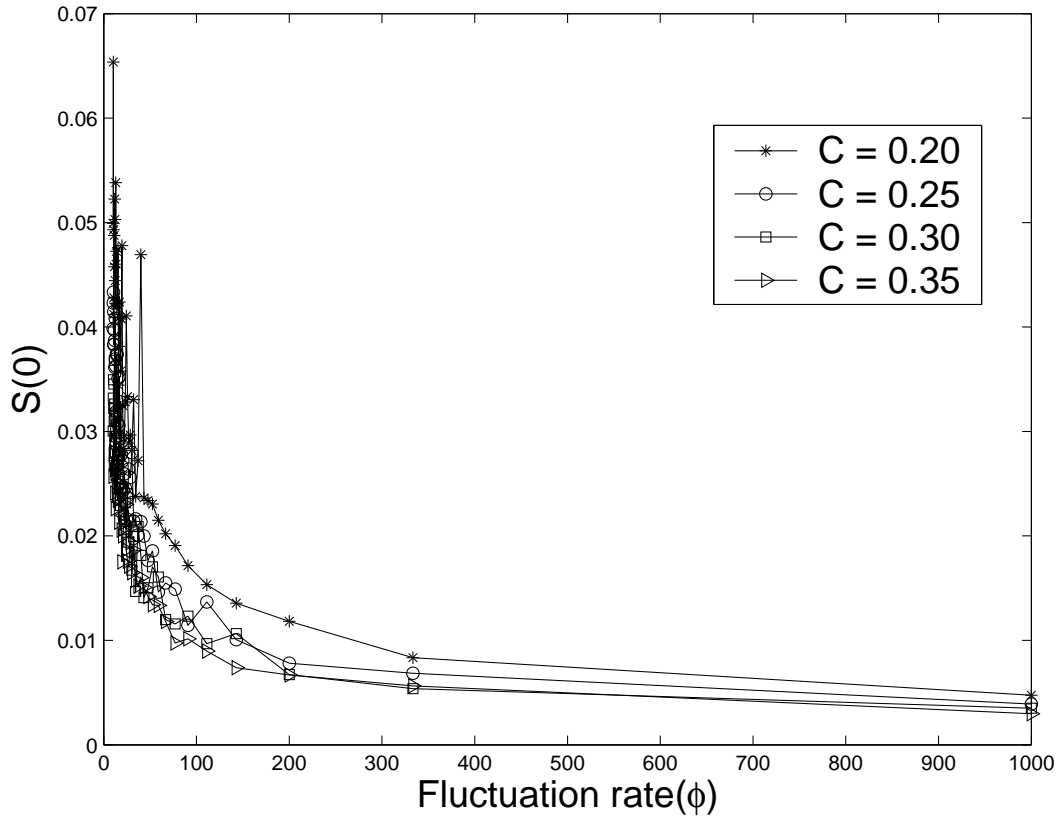}
\caption{The synchronization error decreases with the increase in
the fluctuation rate. It can be seen that high coupling could not
stabilize synchronization with lower fluctuation rates. Here
$\widetilde{\Delta p} =0.05$.}
\label{e_rate}       % Give a unique label
\end{figure}

\begin{figure}
\includegraphics[width=0.9\columnwidth]{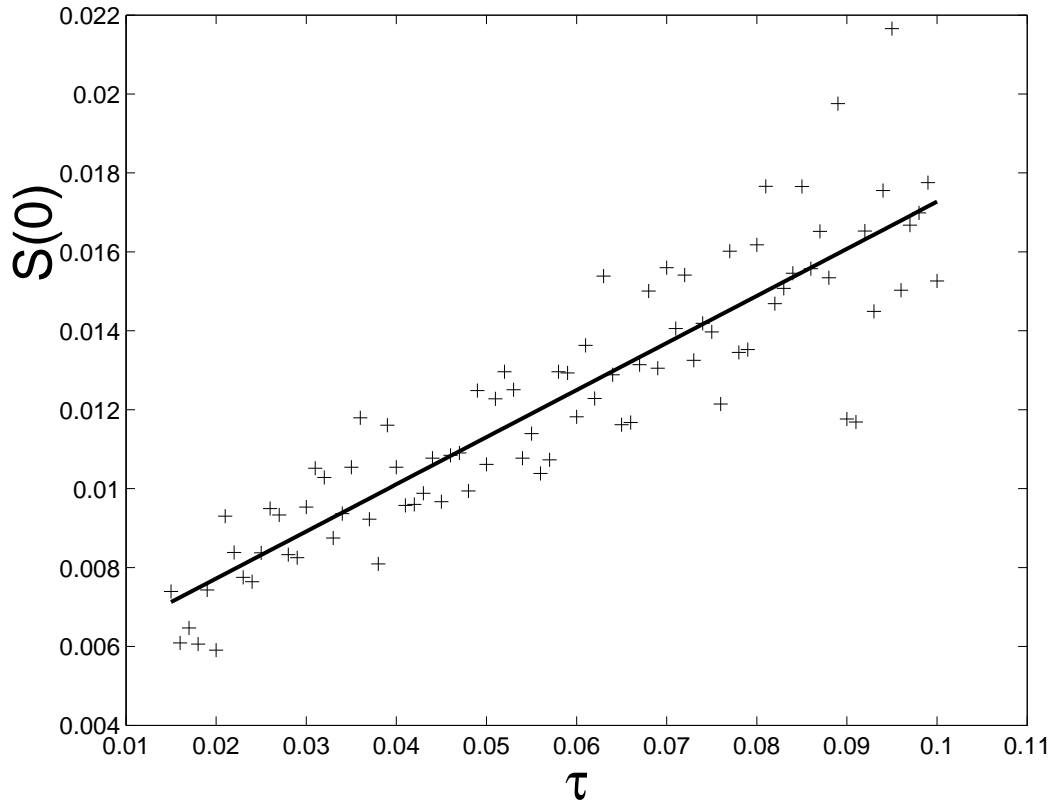}
\caption{Relation between $\phi$ and $S(0)$ is found to be of the
form $S(0)=a+b* \tau_0$, a=0.005 and b= 0.12.}
\label{fit}       % Give a unique label
\end{figure}

\begin{figure}
\includegraphics[width=0.9\columnwidth]{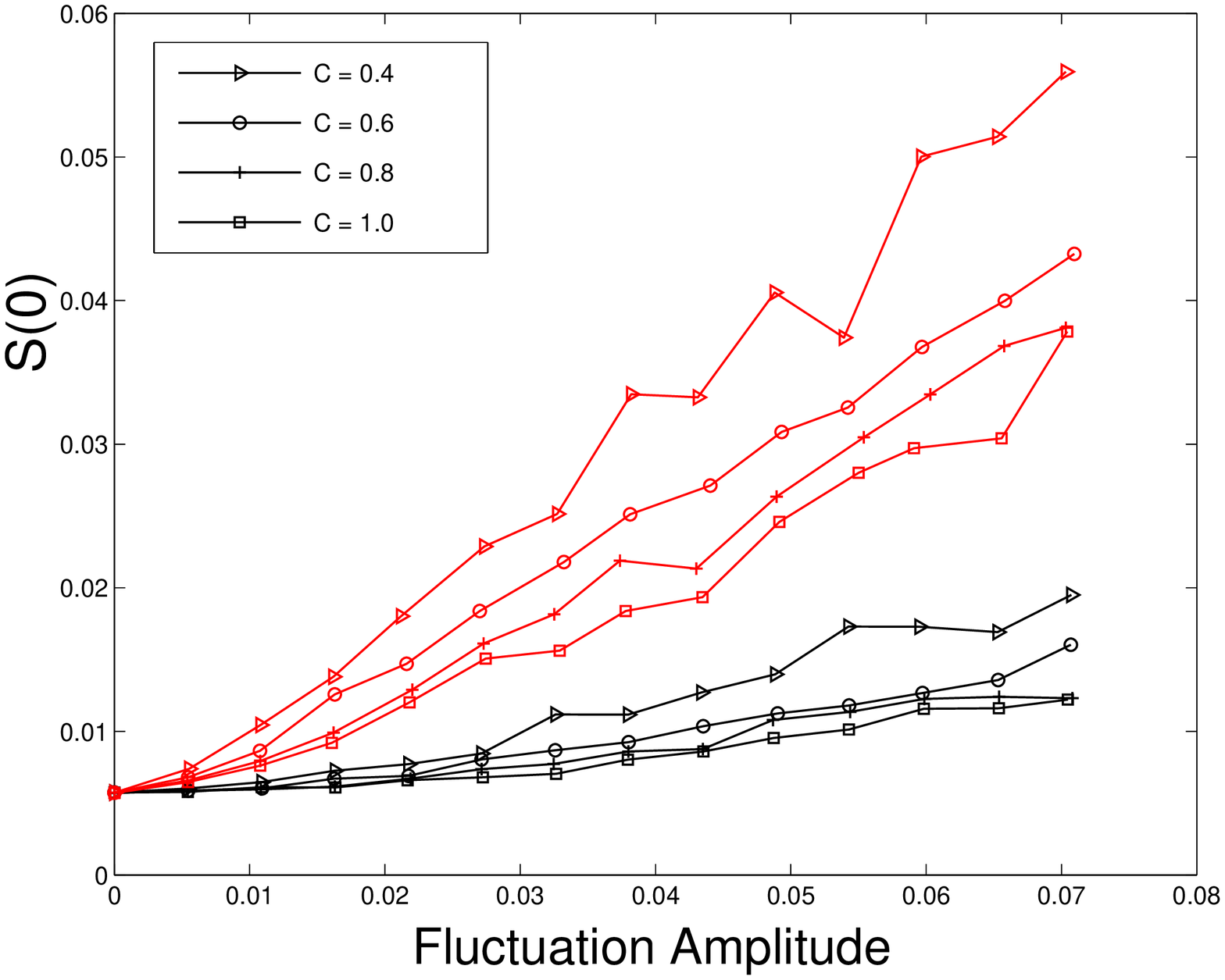}
\caption{For two fluctuation rates, $\phi~=~50$ (red) and $\phi~ =~
500$(black): the synchronization error grows as the amplitude of
fluctuation is increased. Note the less significant role of coupling
strengths compared to that of fluctuation rates.}
\label{growth}       % Give a unique label
\end{figure}

In figure \ref{e_rate} it can be seen that the coupling strength
greater than the threshold value does not play any significant role
in determining the quality of synchronization. But the quality
increase as the fluctuation rate is increased. In figure \ref{fit}
curve fitting is done for the coupling strength $c~=~0.9$, it can be
seen that $S(0)$ varies with $\tau_0$ as
\begin{equation}
S(0) =a+b \tau_0.
\end{equation}
where $a=0.005$ and $b=0.12$ In figure \ref{growth} it is shown that
for two fluctuation rates, $S(0)$ grows linearly with the amplitude
of fluctuations, the growth rate of the error is higher for a larger
waiting time. Numerically, the results were similar for the averages
of $\xi$ over the waiting time following uniform or Gaussian
distribution. Thus for a given amplitude of fluctuations it is the
time scales associated with the fluctuations that determine the
quality of synchronization. Interestingly there are similar results
in biological systems \cite{xi_bio} with coloured fluctuations that
higher correlation times (time scales) makes the coupled systems
less synchronizable.

With a low fluctuation rate, the parameter fluctuations can
considerably affect synchronization because the phase space
evolution time is comparable to the interval where a fixed parameter
mismatch persists. Thus, the system always get time to respond to
the parameter mismatch before it being canceled out. The
error-timescale relations in this regime can be different, which
requires further work.
\section{Conclusion}
In this paper, we show that the synchronization error introduced due
to a deviation of the parameter from its desired value, is
proportional to the waiting time, the instantaneous value of the
variables and the amplitude of perturbation. Asymptotically this
leads to a relation between the similarity function and the
fluctuation rates which is reciprocal in nature. Also the coupling
strength which plays an important role in determining the nature of
synchronization in systems with constant parameter mismatch, but do
not have any significant role when the mutual parameter mismatch is
fluctuating. It is hoped that this investigation will spur further
research in this field, from a more fundamental point of view as
well as for practical implications where high quality
synchronization is required.
\section{Acknowledgements}
We gratefully acknowledge fruitful discussions of this work with Dr.
S.Rajesh. First two authors are supported by the Council for
Scientific and Industrial Research (CSIR), New Delhi.


\begin{thebibliography}{21}
\bibitem{peco1}{L. M. Pecora and T. L. Carroll 1990 Phys. Rev. Lett \textbf{64} 821}
\bibitem{peco2}{T. L. Carroll and L. M. Pecora 1991 IEEE Trans. Circuits Syst.
\textbf{38} II 453}
\bibitem{yamada}{T. Yamada and H. Fujisaka 1983 Prog. Theor. Phys. \textbf{70} 1240}
\bibitem{war1}{G. D. Van Wiggeren and R. Roy 1994 Science \textbf{279}
1198.}
\bibitem{war2}{G. D. Van Wiggeren and R. Roy 1998 Phys. Rev. Lett.\textbf{81}
3547}
\bibitem{roy2}{P. Colet and R. Roy 1994 Opt. Lett. \textbf{19} 2056}
\bibitem{bindu}{V. Bindu and V. M. Nandakumaran 2002 J. Opt. A: Pure Appl. Opt.
\textbf{4} 115}
\bibitem{yin}{H. W. Yin and J. H. Dai and H. J. Zhang 1998 Phys. Rev. E \textbf{58} 9683}
\bibitem{yin}{H. W. Yin and J. H. Dai and H. J. Zhang, Phys. Rev. E \textbf{58}
9683}
\bibitem{phase}{M. G. Rosenblum and A. S. Pikovsky and J. Kurths 1995 Phys. Rev. Lett.
\textbf{76}1804}
\bibitem{pslag}{M. G. Rosenblum A. S. Pikovsky and J. Kurths 1997 Phys. Rev. Lett.
\textbf{78} 4193}
\bibitem{kurth}{Synchronization: A universal concept in nonlinear sciences,
Cambridge University Press, Cambridge 2001}
\bibitem{osc_ein}{The Noisy oscillator, the first hundread years, from Einstein untill Now,
World Scientific, Singapore 2005}
\bibitem{funda}{L. M. Pecora and T. L. Carroll and G. A. Johnson 1997 and D. J. Mar,
Chaos \textbf{7} 520}
\bibitem{xi_bio}{Jacques Rougemont and Felix Naef, Mol. Syst. Biol.\textbf{3} 93}

\end{thebibliography}
\end{document}